\documentclass[preprintnumbers,superscriptaddress,showkeys,showpacs,
byrevtex]{revtex4-1}
\usepackage{amsmath,amsfonts,amssymb,amscd,amsxtra,amsthm}
\usepackage{graphicx}
\usepackage{epstopdf}
\usepackage{bm}
\begin{document}
\preprint{INHA-NTG-03/2019}
\title{$K^0\Lambda$ Photoproduction with Nucleon Resonances}
\author{Sang-Ho Kim}
\email[E-mail: ]{sangho\_kim@korea.ac.kr}
\affiliation{Center for Extreme Nuclear Matters (CENuM),
Korea University, Seoul 02841, Republic of Korea}
\affiliation{Department of Physics, Pukyong National University (PKNU),
Busan 48513, Republic of Korea}
\author{Hyun-Chul Kim}
\email[E-mail: ]{hchkim@inha.ac.kr}
\affiliation{Department of Physics, Inha University, Incheon 22212, Korea}
\affiliation{Advanced Science Research Center, Japan Atomic Energy Agency,
Shirakata, Tokai, Ibaraki, 319-1195, Japan}
\affiliation{School of Physics, Korea Institute for Advanced Study (KIAS),
Seoul 02455, Korea}
\date{\today}
\begin{abstract}
We investigate the reaction mechanism of $K^0 \Lambda$ photoproduction off
the neutron target, i.e., $\gamma n \to K^0 \Lambda$, in the range of $W
\approx 1.6-2.2$ GeV. 
We employ an effective Lagrangian method at the tree-level Born approximation
combining with a Regge approach.
As a background, the $K^*$-Reggeon trajectory is taken into account in
the $t$ channel and $\Lambda$ and $\Sigma$ hyperons in the
$u$-channel Feynman diagram. In addition, the role of various nucleon
resonances listed in the Particle Data Group (PDG) is carefully
scrutinized in the $s$ channel where the resonance parameters are
extracted from the experimental data and constituent quark model.
We present our numerical results of the total and differential cross
sections and compare them with the recent CLAS data.
The effect of the narrow nucleon resonance $N(1685,1/2^+)$ on cross
sections is studied in detail and it turns out that its existence is
essential in $K^0 \Lambda$ photoproduction to reproduce the CLAS data.
\end{abstract}
\maketitle
\section{Introduction}
\label{SecI}
The investigation of kaon photoproduction at low energies plays an
important role in understanding the features of baryon resonances
which weakly couple to $\pi N$ and $\eta N$ channels.
There have been a lot of theoretical and experimental works on the
charged $K^+ \Lambda$ photoproduction in the literature, e.g., 
MAID~\cite{Drechsel:2007if} and SAID~\cite{Arndt:2006bf}, and the
contributions of various nucleon resonances are examined.
In the neutral $K^0 \Lambda$ channel, the background
contribution is much reduced, compared to its charged channel due to
the absence of charged particle exchanges.
Moreover,it is related to the narrow nucleon resonance
$N(1685,1/2^+)$, of which the existence is previously supported in the
$\gamma n \to \eta n$ reaction revealing a clear bump structure on the
cross section near $W = 1.68$ GeV~\cite{Kuznetsov:2006kt}.
This phenomena is called the \textit{neutron anomaly} in $\eta$
photoproduction. It is expected that the $N(1685,1/2^+)$ could
be observed also in $K^0 \Lambda$ photoproduction, since its
threshold is below the mass of this narrow resonance and the $s \bar
s$ component is contained in the $N(1685,1/2^+)$.
In the present talk, we summarize a recent work~\cite{Kim:2018qfu},
where  $K^0 \Lambda$ photoproduction was studied with emphasis on the 
$N(1685,1/2^+)$. 
\section{Theoretical Formalism}
\label{SecII}
\begin{figure}[h]
\centering
\includegraphics[width=9.5cm]{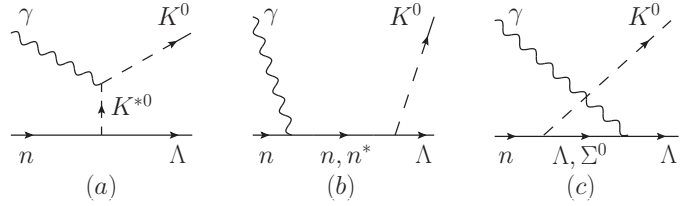}
\caption{Tree-level Feynman diagrams for the $\gamma n \to K^0
  \Lambda$, which consist of (a) $t$-channel $K^*$ Reggeon exchange, 
(b) $s$-channel nucleon and nucleon resonances, and
(c) ground state hyperons in the $u$ channel.}  
\label{fig:1}
\end{figure}

We employ an effective Lagrangian method at the tree-level Born
approximation to study the reaction mechanism of $\gamma n \to K^o
\Lambda$. The relevant Feynman diagrams are drawn in
Fig.~\ref{fig:1}. 
We refer to Ref.~\cite{Kim:2018qfu} for detailed explanation of the
theoretical formalism of the background term and the relevant form
factors. In this talk, we concentrate on the structures of effective
Lagrangians and coupling constants for the resonance term as shown in 
Fig.~\ref{fig:1}(b). The effective Lagrangians for $K \Lambda N^*$
interactions are written as  
\begin{align}
\label{eq:ResLag}
\mathcal{L}^{1/2^\pm}_{K \Lambda N^*}&= - i g_{K \Lambda N^*} \bar K
                                       \bar \Lambda \Gamma^{(\pm)} N^*
                                       + \mathrm{h.c.}, 
\hspace{2.7em}
\mathcal{L}^{3/2^\pm}_{K \Lambda N^*} =
\frac{g_{K \Lambda N^*}}{M_K} \partial^\mu \bar K \bar \Lambda
                                       \Gamma^{(\mp)} 
N^*_\mu + \mathrm{h.c.},
\cr
\mathcal{L}^{5/2^\pm}_{K \Lambda N^*} &=
\frac{ig_{K \Lambda N^*}}{M_K^2} \partial^\mu \partial^\nu \bar K \bar \Lambda
\Gamma^{(\pm)} N^*_{\mu\nu} + \mathrm{h.c.},
\hspace{1em}
\mathcal{L}^{7/2^\pm}_{K \Lambda N^*} =
- \frac{g_{K \Lambda N^*}}{M_K^3} \partial^\mu \partial^\nu \partial^\alpha
\bar K  \bar \Lambda \Gamma^{(\mp)} N^*_{\mu\nu\alpha} + \mathrm{h.c.},
\end{align}
where the $\Gamma^+ (\Gamma^-) = \gamma^5 (1)$ is determined according to
the parity of the nucleon resonances.
$N^*$, $N^*_\mu$, $N^*_{\mu\nu}$, and $N^*_{\mu\nu\alpha}$ are the
Rarita-Schwinger spin $J=$ $\frac{1}{2}$, $\frac{3}{2}$, $\frac{5}{2}$, and
$\frac{7}{2}$ fields, respectively~\cite{Rarita:1941mf}.
\begin{table}[t]
\caption{Nucleon resonances listed in the PDG~\cite{Tanabashi:2018oca} and
$N(1685,1/2^+)$ and their strong coupling constants $g_{K \Lambda N^*}$.
The decay amplitudes $G(\ell)$ are obtained from Ref.~\cite{Capstick:1998uh}
and the branching ratio $\mathrm{Br}_{N^* \to K\Lambda}$ from the
PDG~\cite{Tanabashi:2018oca}.}
\label{TAB1}
\begin{tabular}{c|c||cc|cc|c}
\hline
State&Rating&$G(\ell)\,[\mathrm{MeV}^{1/2}]$&$g_{K \Lambda N^*}$&
$\mathrm{Br}_{N^* \to K\Lambda}\,[\%]$&$|g_{K \Lambda N^*}|$&
$g_{K \Lambda N^*}$(final) \\
\hline
$N(1650,1/2^-)$&****
&$-3.3 \pm 1.0$&$-0.78$&$5-15$
&$ 0.59-1.02$&$-0.78$ \\
$N(1675,5/2^-)$&****
&$0.4 \pm 0.3$&$1.23$& & &$1.23$ \\
$N(1680,5/2^+)$&****
&$0.1 \pm 0.1$&$-2.84$& & &$-2.84$ \\
$N(1700,3/2^-)$&***
&$-0.4 \pm 0.3$&$2.34$& & & $2.34$ \\
$N(1710,1/2^+)$&****
&$4.7 \pm 3.7$&$-7.49$&$5-25$&$ 4.2-9.4$&$-4.2$ \\
$N(1720,3/2^+)$&****
&$-3.2 \pm 1.8$&$-1.80$&$4-5$&$1.8-2.0$&$-1.1$ \\
$N(1860,5/2^+)$&**
&$-0.5 \pm 0.3$&$1.40$&seen& &$1.40$ \\
$N(1875,3/2^-)$&***
&$1.7 \pm 1.0$&$-2.47$&seen& &$-2.47$ \\
$N(1880,1/2^+)$&***
&$$&$ $&$12-28$&$4.5-6.4$&$3.0$ \\
$N(1895,1/2^-)$&****
&$2.3 \pm 2.7$&$0.34$&$13-23$&$0.58-0.77$ &$0.34$ \\
$N(1900,3/2^+)$&****
&$$&$ $&$2-20$&$0.53-1.7$&$0.6$ \\
$N(1990,7/2^+)$&**
&$1.5 \pm 2.4$&$0.61$& & &$0.61$ \\
$N(2000,5/2^+)$&**
&$-0.5 \pm 0.3$&$0.61$& & &$0.61$ \\
$N(2060,5/2^-)$&***
&$-2.2 \pm 1.0$&$-0.52$& seen& &$-0.52$ \\
$N(2120,3/2^-)$&***
&$1.7 \pm 1.0$&$-1.05$& & &$-1.05$ \\
$N(2190,7/2^-)$&****
&$-1.1$&$0.67$& & &$0.67$ \\
\hline
$N(1685,1/2^+)$
&$ $&$ $& & &$-0.9$ \\
\hline
\end{tabular}
\end{table}

We derive the strong coupling constants, $g_{K\Lambda N^*}$, in
Eq.~(\ref{eq:ResLag}) from the quark model predictions where
information on the partial decay amplitude for the $N^* \to K \Lambda$
channel is provided~\cite{Capstick:1998uh}. They have the
relation~\cite{Kim:2017nxg}: 
\begin{align}
\label{eq:DA}
\langle K(\vec{q})\,\Lambda(-\vec{q},m_f) | -i \mathcal{H}_\mathrm{int} |
N^*({\vec 0},m_j) \rangle
=4 \pi M_{N^*} \sqrt{\frac{2}{|\vec{q}|}}  \sum_{\ell,m_\ell}
\langle \ell\, m_\ell\, {\textstyle\frac{1}{2}}\, m_f | j \,m_j \rangle 
Y_{\ell,m_\ell} ({\hat q}) G(\ell),
\end{align}
where $\langle\ell\,m_\ell\,\frac{1}{2}\,m_f|j \,m_j \rangle$ and
$Y_{\ell,m_\ell} ({\hat q})$ are the Clebsch-Gordan coefficients and
spherical harmonics, respectively.
Then, the decay width is calculated from the partial decay amplitude
$G(\ell)$
\begin{align}
\label{eq:DA2}
\Gamma_{N^* \to K \Lambda} = \sum_\ell |G(\ell)|^2 .
\end{align}
We introduce sixteen nucleon resonances from the PDG~\cite{Tanabashi:2018oca}
that are likely to couple strongly to the $\gamma N$ and $K \Lambda$
vertices and the $N(1685,1/2^+)$. All the relevant parameters are listed in
Table~\ref{TAB1}. Here we compare the results from the quark model
predictions with those from the experimental data.
The values of the coupling constants are mostly close to each other, 
although the signs can be determined only from the quark model. 
In the case of the $N(1880,1/2^+)$ and $N(1900,3/2^+)$, there exist
only the experimental data. Thus their strong coupling constants are
determined by using the PDG data and their signs are determined
phenomenologically. 

\section{Results and  Discussions}
\label{SecIII}
\begin{figure}[tbh]
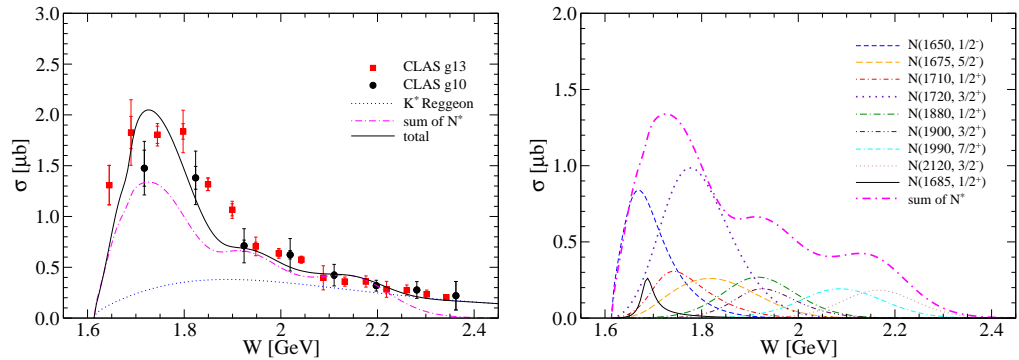

\vspace{0.8em}
\centering
\includegraphics[width=6.5cm]{fig2a.eps} \,\,\,
\includegraphics[width=6.5cm]{fig2b.eps}
\caption{Left: Total cross section for the $\gamma n \to K^0 \Lambda$
as a function of $W$. Right: Separate contribution of various nucleon
resonances to the $\gamma n \to K^0 \Lambda$.
The data are from the CLAS Collaboration~\cite{Compton:2017xkt}.}
\label{fig:2}
\end{figure}

We present the results of our calculation for the total cross section in
Fig.~\ref{fig:2}. It is shown that the nucleon resonances play a
crucial role in reproducing the CLAS data~\cite{Compton:2017xkt}.
At the threshold region (W $\leq$ 1.85 GeV), the contribution of
$K^*$-Reggeon exchange (dotted line) reaches only the level of around
$25\,\%$ compared with CLAS data. The main contributions of the
nucleon resonances are plotted in the right panel of Fig.~\ref{fig:2}.
The $N(1650,1/2^-)$ and $N(1720,3/2^+)$ are mostly responsible
and the magnitude from the narrow resonance $N(1685,1/2^+)$ reaches
around 0.25 $\mathrm{\mu b}$.

\begin{figure}[t]
\centering
\includegraphics[width=13cm]{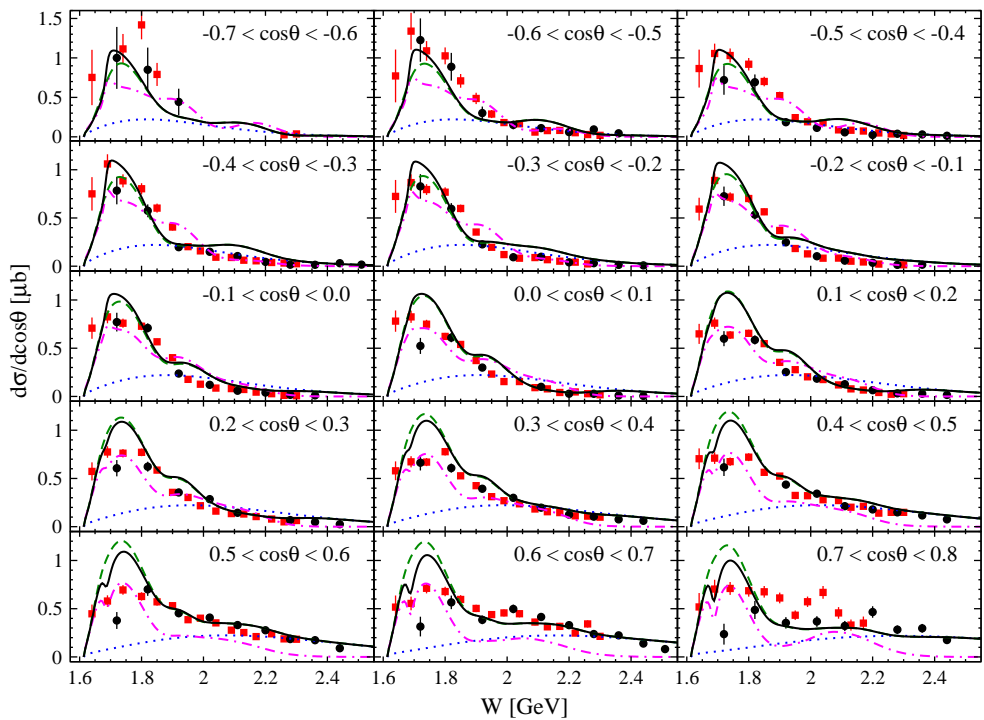}
\caption{Differential cross sections for the $\gamma n \to K^0 \Lambda$
as a function of $W$ at different scattering angles. 
The data are from the CLAS Collaboration~\cite{Compton:2017xkt}.} 
\label{fig:3}
\end{figure}

Figure~\ref{fig:3} draws the numerical results of the differential cross
sections. The dotted, dot-dashed, and solid curves indicate the
contribution from $K^*$-Reggeon exchange, that from the $N^*$
exchanges, and the total contribution, respectively.
The dashed one stands for the total contribution except for the narrow
resonance $N(1685,1/2^+)$. The inclusion of the $N(1685,1/2^+)$
enhances the results in the backward angle and reduces them in the
forward angle by the interference effect. It is also interesting that
the dip structures begin to appear as the angle $\cos\theta$
increases, which is consistent with the present results. This
indicates strongly the existence of the $N(1685,1/2^+)$ in the $\gamma
n \to K^0 \Lambda$. 
\section{Summary and conclusion}
\label{SecIV}
We investigated the mechanism of the $K^0 \Lambda$ photoproduction
with the help of the effective Lagrangian method. We have found that
the $N(1650,1/2^-)$ and $N(1720,3/2^+)$ give the most 
dominant contributions to the cross section. While the effect of the 
$N(1685,1/2^+)$ on the total cross section is very small, it
is essential to describe the differential cross sections near the
threshold region (1.6 $\leq$ W $\leq$ 1.8 GeV).

\section*{Acknowledgment}
This work was supported by the National Research Foundation of Korea
(NRF) grant funded by the Korea government(MSIT)
(No. 2018R1A5A1025563). 

\end{document}